\documentclass[a4paper]{aa}
\usepackage{graphicx}
\usepackage{txfonts}
\usepackage{color}
\usepackage{natbib}
\bibpunct{(}{)}{;}{a}{}{,} 

\begin{document}


\title{Atmospheric constraints for the CO$_2$ partial pressure on terrestrial planets near the outer edge of the habitable zone}
\titlerunning{Constraints on CO$_2$ partial pressure near the outer HZ edge}

\author{P. von Paris\inst{1,2,3}  \and J.L. Grenfell\inst{4} \and P. Hedelt \inst{1,2}\thanks{current address: Institut f\"{u}r Methodik der Fernerkundung,\newline Deutsches Zentrum f\"{u}r Luft-und Raumfahrt (DLR),\newline Oberpfaffenhofen, 82234 We{\ss}ling, Germany}  \and H. Rauer\inst{3,4}  \and F. Selsis\inst{1,2} \and B. Stracke\inst{3}}

\institute{Univ. Bordeaux, LAB, UMR 5804, F-33270, Floirac, France \and CNRS, LAB, UMR 5804, F-33270, Floirac, France 
\and Institut f\"{u}r Planetenforschung, Deutsches Zentrum
f\"{u}r Luft- und Raumfahrt (DLR), Rutherfordstr. 2, 12489 Berlin, Germany
\and  Zentrum f\"{u}r Astronomie und Astrophysik (ZAA), Technische
Universit\"{a}t Berlin, Hardenbergstr. 36, 10623 Berlin, Germany}

\abstract {In recent years, several potentially habitable, probably terrestrial exoplanets and exoplanet candidates have been discovered. The amount of CO$_2$ in their atmosphere is of great importance for surface conditions and habitability. In the absence of detailed information on the geochemistry of the planet, this amount could be considered as a free parameter.}{Up to now, CO$_{2}$ partial pressures for terrestrial planets have been obtained assuming an available volatile reservoir and outgassing scenarios. This study aims at calculating the allowed maximum CO$_{2}$ pressure at the surface of terrestrial exoplanets orbiting near the outer boundary of the habitable zone by coupling the radiative effects of the CO$_2$ and its condensation at the surface. These constraints might limit the permitted amount of atmospheric CO$_{2}$, independent of the planetary reservoir.}{A 1D radiative-convective cloud-free atmospheric model was used to calculate surface conditions for hypothetical terrestrial exoplanets. CO$_{2}$ partial pressures are fixed according to surface temperature and vapor pressure curve. Considered scenarios cover a wide range of parameters, such as gravity, central star type and orbital distance, atmospheric N$_{2}$ content and surface albedo. }{Results show that for planets in the habitable zone around K-, G-, and F-type stars the allowed CO$_2$ pressure is limited by the vapor pressure curve and not by the planetary reservoir. The maximum CO$_2$ pressure lies below the CO$_2$ vapor pressure at the critical point of $p_{\rm{crit}}=$73.8\,bar. For M-type stars, due to the stellar spectrum being shifted to the near-IR, CO$_2$ pressures above $p_{\rm{crit}}$ are possible for almost all scenarios considered across the habitable zone. This implies that determining CO$_{2}$ partial pressures for terrestrial planets by using only geological models is probably too simplified and might over-estimate atmospheric CO$_{2}$ towards the outer edge of the habitable zone.}{}


\maketitle

\section{Introduction}

Given the difficulties and challenges of detecting sub-surface life on Earth, any life to be first discovered
beyond our own solar system will most likely  be restricted to the
planetary surface and atmosphere. This is the basis of the concept of the habitable zone (HZ, e.g., \citealp{dole1964}, \citealp{hart1978}, \citealp{kasting1993}). The HZ is defined as the region around a star where a rocky planet with a suitable atmosphere can host liquid water on its surface, a condition motivated by the fact that all life as we know it requires liquid water.

Several studies have implied that small, potentially rocky planets are common (e.g., \citealp{howard2010_occur}, \citealp{Wittenmyer2011}, \citealp{borucki2011}, \citealp{mayor2011}, \citealp{cassan2012}, \citealp{gaidos2012}). Hence, it is not unreasonable to assume that planets in the HZ of their central stars may also be relatively common. Indeed, some potentially habitable (candidate) super-Earths in or very close to the HZ of their central star have already been discovered (\citealp{udry2007}, \citealp{mayor2009gliese},  \citealp{borucki2011}, \citealp{pepe2011}, \citealp{bonfils2011}, \citealp{anglada2012}, \citealp{delfosse2012}, \citealp{borucki2012}). Also, Neptune- or Jupiter-like planets have been discovered in the HZ (e.g., \citealp{lovis2006}, \citealp{fischer2008}, \citealp{haghighipour2010}, \citealp{tinney2011}) which raises the possibility of habitable satellites around these planets.

A simple criterion for the potential habitability of a planet, which is immediately accessible from the discovery data, is its equilibrium temperature, $T_{\rm{eq}}$. The equilibrium temperature is calculated by

\begin{equation}
\label{eqtemp}
T_{\rm{eq}}=\left(\frac{(1-A)F}{4\sigma}\right)^{0.25}
\end{equation}

where $A$ is the planetary albedo, $F$ the stellar flux at the orbital distance of the planet and $\sigma$ the Stefan-Boltzmann constant. As was discussed by, e.g., \citet{selsis2007gliese} and \citet{kaltenegger2011kepler}, a habitable planet should have $T_{\rm{eq}} \lesssim$\,270\,K to avoid a runaway heating of the surface and corresponding loss of the complete surface water reservoir. For low values of $T_{\rm{eq}}$ near the outer edge of the HZ (e.g., model calculations for GL 581\,d suggest $T_{\rm{eq}}\sim 190$\,K, \citealp{vparis2010gliese}), a massive greenhouse effect must be provided by the atmosphere to obtain habitable surface conditions. 

H$_{2}$O is the most obvious candidate of radiatively active gases which could provide the necessary greenhouse warming. It provides the bulk of the greenhouse effect on Earth. Furthermore, H$_{2}$O is by definition present on the surface of a habitable planet. The H$_{2}$O partial pressure in an atmosphere of a potentially habitable planet is controlled by evaporation (or sublimation) from the surface reservoir, taking into account the water vapor pressure curve. Besides H$_{2}$O, CO$_{2}$ is usually considered the most important greenhouse gas for the determination of the outer boundary of the HZ (e.g., \citealp{kasting1993}). 

On Earth, CO$_{2}$ is controlled by processes such as volcanic outgassing or rock weathering. To estimate CO$_{2}$ partial pressures for terrestrial exoplanets, up to now only geological models were used (e.g., \citealp{elkins2008}, \citealp{kite2009}, \citealp{kite2011}, \citealp{edson2012}, \citealp{abbot2012}). Furthermore, the volatile content of habitable zone planets is expected to be highly variable due to orbital migration (e.g., \citealp{raymond2004}). For instance, planets originating from the outer planetary system and made of a large fraction of cometary material can migrate to habitable orbital distances, resulting in the so-called ocean-planets \citep{leger2004}. Planetary CO$_{2}$ reservoirs of the order of thousands of bars are certainly plausible, when considering typical solar system values for the composition of the cometary material. It is possible that the silicate carbonate cycle, which regulates the level of atmospheric CO$_2$ on Earth, does not operate on ocean planets in the absence of continents. Such large reservoirs of CO$_2$ are therefore a concern for habitability if totally outgassed into a CO$_{2}$-rich envelope.

\begin{figure}[h]
  \resizebox{\hsize}{!}{ \includegraphics*{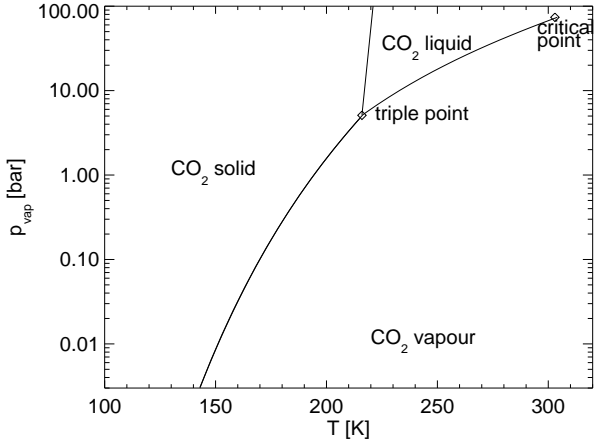}} \\
 \caption{CO$_{2}$ phase diagram.}
 \label{co2phase}
\end{figure}

Fig. \ref{co2phase} shows the phase diagram of CO$_{2}$. The critical point lies at $T_{\rm{crit}}$=303\,K and $p_{\rm{crit}}$=73.8\,bar. At a given surface temperature below $T_{\rm{crit}}$, the vapor pressure curve actually limits the amount of CO$_{2}$ which can be outgassed into the atmosphere, independent of the planetary reservoir. 

It is the aim of this study to quantify this maximum CO$_{2}$ partial pressure for a range of possible planetary scenarios near the outer edge of the HZ, based on the phase diagram in Fig. \ref{co2phase}. In order to put  constraints on atmospheric CO$_{2}$, the interplay between CO$_{2}$ greenhouse effect, surface temperature and CO$_2$ partial pressure must be taken into account. Therefore, this work will use an atmospheric model which consistently calculates temperature profiles and surface conditions. It will be investigated how different parameters such as planetary gravity, orbital distance and central star type, N$_{2}$ pressure and surface albedo influence the behavior of the maximum CO$_2$ partial pressure.

The paper is organized as follows: Sect. \ref{model} presents the model and scenarios, Sect. \ref{results} the results and Sect. \ref{discuss} a discussion. We conclude with Sect. \ref{summary}.

\section{Methods}

\label{model}

\subsection{Atmospheric model}

We used a cloud-free, one-dimensional radiative-convective model to determine the globally averaged atmospheric temperature-pressure profile.

The original model was first described by
\citet{kasting1984water} and \citet{kasting1984}. Further
developments were introduced by e.g. \citet{kasting1988}, \citet{kasting1991}, \citet{kasting1993}
\citet{mischna2000}, \citet{pavlov2000} and \citet{Seg2003}. The model version used in this work
is taken from \citet{vparis2008} and
\citet{vparis2010gliese} where more details on the model are given.

The model atmospheres are assumed to be composed of N$_2$, H$_2$O, and CO$_2$. Temperature profiles are obtained on 52 model layers, approximately spaced equidistantly in $\log$\,(pressure). The pressure grid is determined from the surface pressure $p_{\rm{surf}}$ (variable, see below) up to a pressure of 6.6$\times$10$^{-5}$ bar (fixed) at the model lid.

The model calculates the temperature profile by solving the radiative
transfer equation. The radiative fluxes are calculated separately for the stellar (mostly visible) and the planetary (mostly IR) flux. The stellar part of the radiative transfer uses gaseous opacities from \citet{pavlov2000} and Rayleigh scattering formulations from \citet{vparis2010gliese}. Gaseous opacities in the IR are based on Hitemp data \citep{rothman1995} and continuum absorption adapted from \citet{clough1989} and \citet{kasting1984water}. The purpose of the 1D model used here is to calculate an arbitrary range of temperature-pressure scenarios ranging from the outer to the inner boundary of the HZ. Therefore, we used Hitemp in order to have reliable results for wet, hot atmospheres. The choice of the specific opacity database (e.g., Hitran 2008, Hitran 2004, etc.) for gaseous absorption is not critical for the results presented below, i.e. relatively dry, cold scenarios.

If the calculated radiative lapse rate is sub-adiabatic, the model performs convective adjustment, assuming a wet adiabatic lapse rate. This wet adiabatic lapse rate is determined considering either CO$_2$ or H$_2$O as condensing species.

The treatment of CO$_2$ condensation for the calculation of the adiabatic lapse rate follows \citet{vparis2010gliese}. We assume that CO$_2$ condensation occurs when the atmosphere is super-saturated with
respect to CO$_2$, as described by the super
saturation ratio $\mathcal{S}_{\rm{s}}$:

\begin{equation}\label{co2glan}
    \frac{p_{\rm{CO_2}}}{p_{\rm{vap},CO_2}} = \mathcal{S}_{\rm{s}} = 1.34
\end{equation}

where $p_{\rm{CO_2}}$ is the partial CO$_2$ pressure and
$p_{\rm{vap},CO_2}$ the saturation vapor pressure of CO$_2$. The
chosen value of $\mathcal{S}_{\rm{s}}$ is motivated by measurements reported in \citet{glandorf2002}. Condensation of an atmospheric constituent can occur when $\mathcal{S}_{\rm{s}}$ is closer to unity than the value chosen here. Note that other studies (e.g.,  \citealp{kasting1991} or \citealp{kasting1993}) assumed $\mathcal{S}_{\rm{s}}$=1 which represents the thermodynamic lower limit where condensation could occur.

The water profile in the model is calculated based on the relative humidity distribution of \citet{manabewetherald1967}. Above the cold trap,
the water profile is set to an isoprofile taken from the cold trap value. Despite the fact that CO$_2$ is allowed to condense, the major atmospheric constituents N$_{2}$ and CO$_{2}$ are isoprofiles throughout the entire atmosphere, i.e. are assumed to be well-mixed. The impact of fixing the CO$_2$ mixing ratio at the saturation value on the atmospheric energy budget is expected to be rather small, hence would not change our results by much. A more consistent treatment of CO$_2$ condensation (including an altitude-dependent CO$_2$ profile) would involve vertical mass transport and an atmospheric pressure grid which is not in hydrostatic equilibrium in the region of CO$_2$ condensation. Introducing this into our atmospheric model is beyond the scope of the current work.

\subsection{Model procedure}

\begin{table*}
  \centering
  \caption{Parameter (range) for the runs performed}\label{planpar}
  \begin{tabular}{lccccc}
     \hline
   \hline
    Runs &  Gravity [$g_{\oplus}$]&Stellar type& Stellar insolation $S$ & surface albedo $A_{S}$& $p_{N_{2}}$ [bar]   \\
    \hline
    \hline
    nominal             & 1-3 & M-F & 0.2-0.5  &0.13        &1.0\\\
    N$_{2}$ study  & 1     & M-F  & 0.2-0.5 &0.13        &0.1-10.\\\
    $A_{S}$ study  & 1 & M-F & 0.2-0.4  &0.13-0.4 &1.0\\\
          \end{tabular}
\end{table*}

The simulations started with a CO$_2$ partial pressure of 73.8\,bar, corresponding to the pressure at the critical point, $p_{\rm{crit}}$, and an isothermal temperature profile of 320\,K, i.e. higher than the critical temperature of 303\,K. The choice of the initial temperature profile is not critical for the final outcome of the simulations. We did not allow for CO$_2$ partial pressures higher than 73.8\,bar, even though higher pressures are certainly possible (e.g., Venus).

The surface pressure in model iteration step $t+1$ is re-calculated based on the surface temperature $T_{\rm{surf}}$ as

\begin{equation}\label{dynamicpress}
 p_{\rm{surf}}(T_{\rm{surf}})=p_{\rm{N_{2}}}+p_{\rm{H_2O}}(T_{\rm{surf}})+p_{\rm{CO_2}}(T_{\rm{surf}})
\end{equation}

where $p_{\rm{N_{2}}}$ is the fixed background pressure of N$_{2}$. The water vapor pressure is obtained from

\begin{equation}\label{waterpress}
 p_{\rm{H_2O}}(T_{\rm{surf}})=\rm{min}\left(p_{\rm{vap,H_2O}}(T_{\rm{surf}}),p_{\rm{ocean}}\right)
\end{equation}

with $p_{\rm{vap,H_2O}}(T_{\rm{surf}})$ the water vapor saturation pressure at
surface temperature and $p_{\rm{ocean}}$ the ocean reservoir assumed (here, 1 Earth ocean, i.e. 270\,bar). The CO$_2$ partial pressure is accordingly
calculated as

\begin{equation}\label{co2press}
 p_{\rm{CO_2}}(T_{\rm{surf}})=\rm{min}\left(p_{\rm{vap,CO_2}}(T_{\rm{surf}}),p_{\rm{CO_2}}\right)
\end{equation}

Note that this corresponds to assuming a super-saturation ratio of $\mathcal{S}_{\rm{s}}$=1 at the surface, in contrast to $\mathcal{S}_{\rm{s}}$=1.34 used for the atmospheric CO$_{2}$ adiabatic lapse rate (see eq. \ref{co2glan}). This is motivated by the fact that atmospheric condensation generally requires $\mathcal{S}_{\rm{s}}$$>$1 (i.e., the presence of condensation nuclei). At the surface, however, atmosphere and reservoir are in equilibrium, hence the partial pressure follows the vapor pressure curve.

The mixing ratio of N$_{2}$ is then adjusted via

\begin{equation}\label{dynamicpress_vmr}
 C_{N_{2},t+1}=C_{N_{2},t}\cdot \frac{p_{\rm{surf},t}}{p_{\rm{surf},t+1}}
\end{equation}

where $C_{N_{2},t+1}$, $C_{N_{2},t}$ are the N$_{2}$ concentrations and
$p_{\rm{surf},t+1}$, $p_{\rm{surf},t}$ the surface pressures at iteration
steps $(t+1)$ and $t$.

Based on the new value for the surface pressure $p_{\rm{surf}}$, the pressure grid on the 52 model levels is then re-calculated.

\begin{figure}[h]
    \resizebox{\hsize}{!}{\includegraphics*{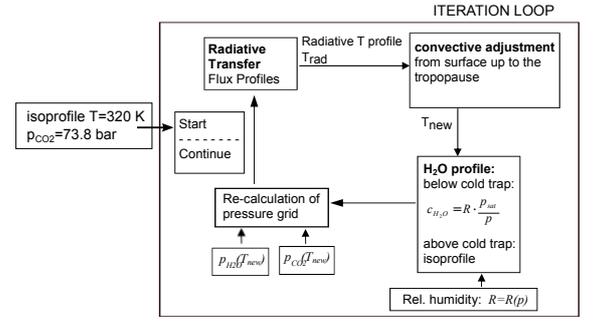}}\\
  \caption{ Flow chart of the model.}
  \label{modeldiagram}
\end{figure}

Fig. \ref{modeldiagram} shows a flow chart of the model to illustrate the model procedure.

The CO$_2$ saturation vapor pressure $p_{\rm{vap,CO_2}}$ is taken from \citet{ambrose1956}. It is divided into two temperature regimes. For $T>216.6$ K (gas over liquid):

\begin{eqnarray}\label{gradco2wethigh}
 \frac{d \ln(p_{\rm{vap,CO_2}})}{d \ln(T)}=&&2.303 \cdot T \cdot\\
\nonumber&& ( \frac{867.2124}{T^2}+18.65612 \cdot10^{-3}- \\
\nonumber &&2\cdot
72.4882\cdot 10^{-6}\cdot T\\\nonumber&& + 3\cdot 93\cdot 10^{-9}T^2)
\end{eqnarray}

For $T\leq 216.6$ K (gas over solid):

\begin{eqnarray}\label{gradco2wethigh}
 \frac{d \ln(p_{\rm{vap,CO_2}})}{d \ln(T)}=&&2.303 \cdot T \cdot\\
\nonumber&&\left( \frac{1284.07}{(T-4.718)^2}+1.256\cdot 10^{-4}\right)
\end{eqnarray}

If surface temperatures remain above 303\,K throughout the entire simulation, the maximum CO$_{2}$ partial pressure is assumed to lie above the critical pressure. However, if surface temperatures converge to values below 303\,K, the corresponding CO$_{2}$ partial pressure is taken as the maximum possible CO$_{2}$ pressure for the particular planetary scenario.

\subsection{Parameter variations}

\label{para}

We varied five important model parameters: The planetary gravity, related to its mass and radius, the type of the central star and the energy input from the star, related to orbital distance, as well as model surface albedo and N$_{2}$ partial pressure. Table \ref{planpar} summarizes the varied parameters.

\begin{itemize}
  \item We assumed three different values for planetary gravity (1x, 2x and 3x Earth's gravity) which roughly corresponds to planetary masses of 1, 5 and 11 Earth masses, respectively, according to mass-radius relationships for rocky planets (e.g., \citealp{sotin2007}).

  \item We used spectra of AD Leo, $\epsilon$ Eri, the Sun and $\sigma$ Boo as examples for M-, K-, G- and F-type stars, respectively. The same sample of stars has been used for numerous studies regarding the influence of stellar type on atmospheric conditions (e.g., \citealp{Seg2003}, \citealp{Seg2005}, \citealp{Grenf2007asbio}, \citealp{Grenf2007pss}, \citealp{kitzmann2010}). Stellar effective temperatures increased from M- to F-type stars, from about 3,400\,K to 6,700\,K, respectively. A more detailed description of the stellar spectra as well as data sources and references can be found in \citet{kitzmann2010}.
  \item The incoming stellar insolation $S_I$ at the top of the model atmospheres is calculated from

\begin{equation}\label{eninput}
 S_I=S\cdot S_0
\end{equation}

 where $S_0$ is the flux currently received by modern Earth (i.e., $S_0$=1366\,Wm$^{-2}$) and $S$ is a constant factor related to orbital distance (e.g., for Earth, $S$=1). In this study, $S$ was varied from $S$=0.2 to $S$=0.5. Corresponding orbital distances ranged from 0.21-0.34\,AU, 0.85-1.35\,AU, 1.41-2.23\,AU and 2.67-4.22\,AU for the M-, K-, G- and F-type stars, respectively (based on \citealp{kitzmann2010}). The range of stellar insolation considered here roughly covers the outer limit of the HZ for the stellar types used in this work (e.g., GL 581\,d with $S$=0.29 and early Mars with $S$=0.32, are both potentially habitable) as well as orbits slightly closer to or slightly farther away from the central star.
 \item The above runs (nominal runs in Table \ref{planpar})  were performed with the N$_{2}$ partial pressure fixed at 1\,bar.  Increasing the amount of N$_2$ (at fixed values of CO$_{2}$ partial pressures) leads to two competing effects, a cooling effect (related to enhanced Rayleigh scattering), and a warming effect (due to pressure broadening of absorption lines and continuum absorption). Several studies have shown that increasing N$_2$ partial pressures might indeed help to obtain habitable surface conditions in atmospheric simulations (e.g., \citealp{goldblatt2009faintyoungsun}, \citealp{vparis2010gliese}).  Hence, we varied the N$_{2}$ partial pressure from 0.1 to 10\,bar, for the 1$g$ runs (N$_{2}$ study in Table \ref{planpar}). 

 \item For all the model scenarios described above, the measured mean surface albedo of the Earth ($A_{\rm{surf}}$=0.13, taken from \citealp{rossow1999}) is used. However, the surface albedo has an important impact on the calculated surface temperature (e.g., \citealp{vparis2008}, \citealp{rosing2010}, \citealp{wordsworth2010}).  Our model calculations do not take into account the possible increase of surface albedo due to condensing and freezing CO$_2$ during the iterations. In this regard, our calculated CO$_2$ partial pressures are likely to be upper limits. Measurements and modeling of the albedo of CO$_2$ snow by \citet{warren1990} suggest that the albedo of CO$_{2}$ snow and ice might be significantly higher than 0.13. Therefore, we performed additional calculations ($A_{S}$ study in Table \ref{planpar}) with a surface albedo of $A_{\rm{surf}}$=0.4 for the 1\,g scenarios, at stellar insolations corresponding to $S$=0.2 and $S$=0.4 and a N$_{2}$ partial pressure of 1\,bar, respectively. 

\end{itemize}

\begin{figure*}
  \resizebox{\hsize}{!}{ \includegraphics{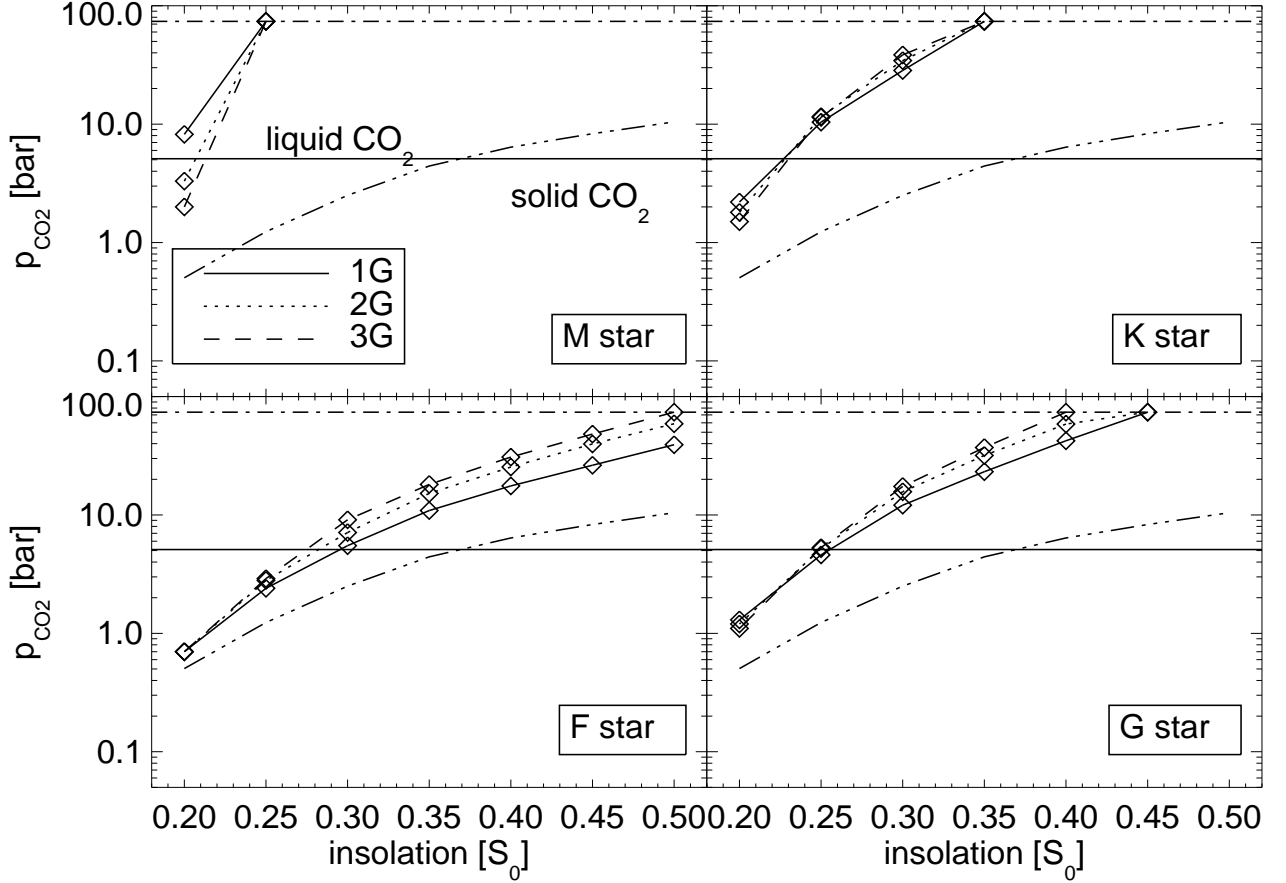}}\\
  \caption{Maximum CO$_2$ partial pressure as a function of gravity and stellar insolation (as defined by Eq. \ref{eninput}). The critical pressure $p_{\rm{crit}}$ is indicated by the dot-dashed horizontal line.  The triple-dot dashed line indicates the highest CO$_2$ pressure calculated for the maximum equilibrium temperature ($A$=0, eq. \ref{eqtemp}).}
  \label{maxpco2}
\end{figure*}

\section{Results}

\label{results}

Fig. \ref{maxpco2} shows the maximum partial pressures of CO$_2$ as a function of stellar insolation (hence, orbital distance, see Eq. \ref{eninput}) for the nominal runs of Table \ref{planpar}. Additionally shown as triple dot-dashed line in Fig. \ref{maxpco2} is the CO$_{2}$ partial pressure when using an equilibrium temperature assuming zero albedo (i.e., the maximum equilibrium temperature, $T_{\rm{eq,max}}$, see eq. \ref{eqtemp}). This shows that detailed atmospheric modeling (taking into account the greenhouse effect) is indeed needed to obtain consistent constraints on the CO$_{2}$ partial pressure. Also indicated in Fig. \ref{maxpco2} (by the horizontal plain line) is the boundary between liquid and solid phase of surface CO$_{2}$, i.e. the triple point pressure of 5.1\,bar (see the phase diagram, Fig. \ref{co2phase}). For maximum CO$_{2}$ pressures below 5.1\,bar, the atmosphere is in equilibrium with CO$_{2}$ ice, above 5.1\,bar, the formation of (shallow) CO$_{2}$ oceans is suggested.

\begin{figure}[h]
    \resizebox{\hsize}{!}{\includegraphics*{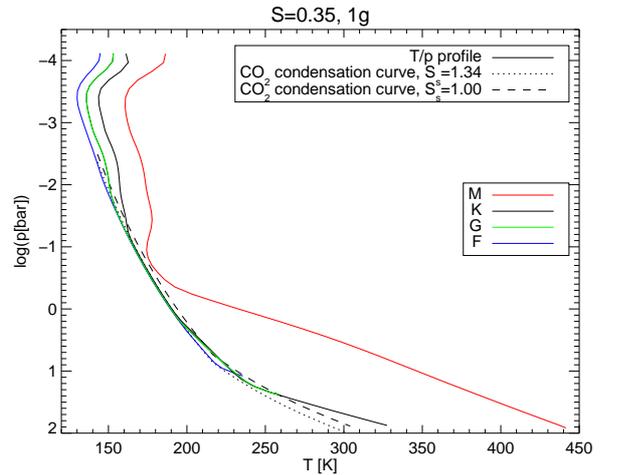}}\\
  \caption{Temperature profile for 1\,g planets at $S$=0.35.}
  \label{samplet}
\end{figure}

Fig. \ref{samplet} shows sample temperatures profile of the simulations, i.e. a 1\,g planet at $S$=0.35, with a N$_{2}$ pressure of 1\,bar and $A_{S}$=0.13. As can be clearly seen, the K- and M-star planets retain their initial CO$_{2}$ inventory of 73.8\,bar (since at the surface, the atmosphere is not saturated with respect to CO$_{2}$), whereas for the F- and G-star planets, CO$_{2}$ partial pressures are below the critical pressure, at 10.9 and 23.2\,bar, respectively. The upper stratosphere is sensitive to absorption of stellar radiation in the near-IR bands of CO$_{2}$ and H$_{2}$O, resulting in about 30\,K increase for an M-star planet compared to the F-star planet. Additionally shown in Fig. \ref{samplet} are the CO$_{2}$ vapor pressure curve ($\mathcal{S}_{\rm{s}}$=1, dashed line, eq. \ref{co2press}) which intersects the temperature profile (for the F star and the G star) at the surface. Furthermore,  Fig. \ref{samplet} shows the CO$_{2}$ condensation curve from eq. \ref{co2glan} ($\mathcal{S}_{\rm{s}}$=1.34) indicating the CO$_{2}$ convective regime. It is clearly seen that the atmospheres of the F-star and the G-star planet are dominated by a CO$_{2}$ convective regime, followed by a very shallow near-surface H$_{2}$O convective regime. In contrast, the K- and M-star planets show a relatively extensive lower troposphere dominated by H$_{2}$O condensation.

\subsection{Effect of stellar type}

\label{stellar_type}

\begin{figure}[h]
    \resizebox{\hsize}{!}{\includegraphics*{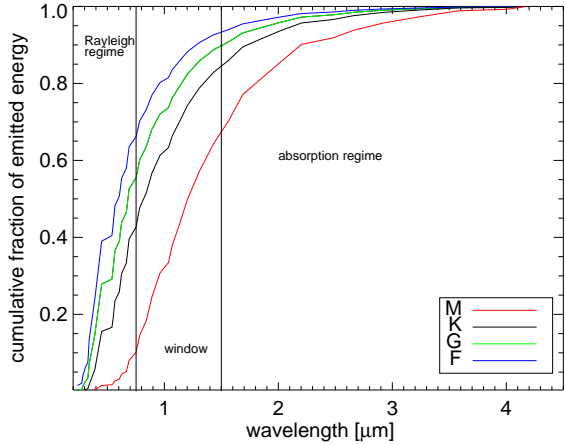}}\\
  \caption{Cumulative energy of different central stars. Regimes are indicated by vertical lines.}
  \label{energybalance}
\end{figure}

From Fig. \ref{maxpco2}, it is clear that with increasing stellar effective temperature (changing stellar type from M to F), the maximum partial pressure of CO$_2$ decreases. Also, the minimum stellar insolation $S_{\rm{min}}$ for which maximum CO$_2$ pressures above $p_{\rm{crit}}$ are possible depends sensitively on the stellar type ($S_{\rm{min}}$=0.25 for the M-star planets and $S_{\rm{min}}\geq$0.5 for the F-star planets). This is due to the distribution of the stellar energy received by the model planets. With increasing stellar effective temperature, the stellar spectrum is shifted towards lower (bluer) wavelengths, as illustrated by Fig. \ref{energybalance}. Broadly, the stellar spectrum can be separated into three regimes, 1) a Rayleigh scattering regime, 2) an absorption regime and 3) a "window" in between. The Rayleigh scattering regime is here defined as the spectral range where the Rayleigh cross section remains larger than 10$^{-2}$ of the maximum value ($\lambda \lesssim $0.75$\mu$m). 

\begin{figure}[h]
 \resizebox{\hsize}{!}{   \includegraphics*{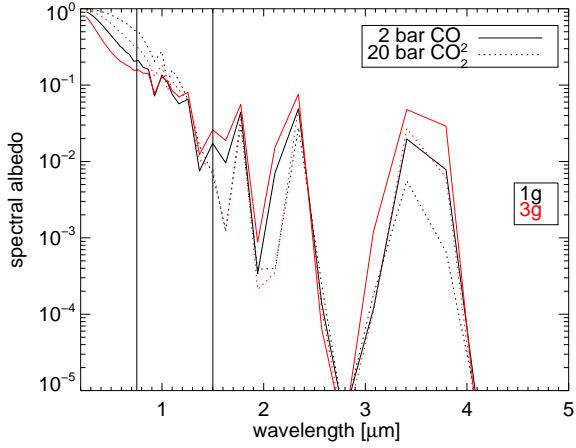}}\\
  \caption{Spectral albedo for a 2 and a 20\,bar CO$_{2}$ atmosphere with surface temperature 288\,K (corresponding to 17\,mbar of H$_{2}$O), 1\,bar of N$_{2}$ and $A_{S}$=0.13. 1\,g and 3\,g planets indicated in black and red, respectively. Window regime is indicated by vertical lines.}
  \label{specalb}
\end{figure}

The absorption regime starts at about 1.5\,$\mu$m where the first strong water and CO$_2$ absorption bands occur. At the high CO$_{2}$ partial pressures considered in this work, both the Rayleigh scattering regime and the absorption regime are almost entirely optically thick to incoming stellar radiation (i.e., no radiation reaching the surface). In the Rayleigh scattering regime, radiation is reflected back to space (high spectral albedo), whereas in the absorption regime, the radiation is deposited in the upper to middle atmosphere (very low spectral albedo), as illustrated in Fig. \ref{specalb} for a 2\, and 20\,bar CO$_{2}$ atmosphere. Depending on spectral type, the actual percentage of stellar radiation contained in the  "window" changes quite considerably, as illustrated in Fig. \ref{energybalance} (around 50\% for the M star, only 30\% for the F star).

Therefore, the planetary albedo becomes larger for increasing stellar effective temperature (M to F) because of the increasingly important contribution of Rayleigh scattering, and thus surface temperatures and corresponding CO$_2$ partial pressures are lower.

\subsection{Effect of planetary gravity}

The most noticeable effect when changing the planetary gravity $g$ is the effect on atmospheric column density $C$. At constant pressure $p$, $C$ and $g$ are related linearly via $C \sim p g^{-1}$. Hence, an increase in gravity leads to a corresponding decrease of atmospheric column density.

This leads to three important effects. Firstly, such a decrease in atmospheric column density leads to decreased Rayleigh scattering, hence a lower planetary albedo (see Fig. \ref{specalb}), hence favors surface warming. Furthermore, less atmospheric column density leads to less near-IR absorption of stellar radiation, hence higher albedo (again, see Fig. \ref{specalb}), hence surface warming (more starlight reaches the surface) and stratospheric cooling. On the other hand, a decreased atmospheric column density leads to less greenhouse effect (GHE), hence surface cooling. The net result on surface temperature when combining these three effects (either cooling or warming) depends on the amount of CO$_{2}$ and the stellar type which determines the planetary albedo and stellar energy distribution (see Figs. \ref{energybalance} and \ref{specalb}).

\begin{figure}[h]
  \resizebox{\hsize}{!}{ \includegraphics*{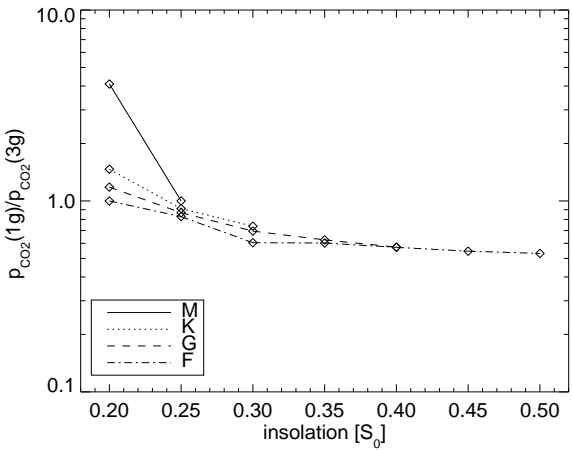}}\\
  \caption{Effect of varying planetary gravity on the calculated maximum CO$_2$ pressures. Ratio between calculated CO$_2$ pressures at 1$g$ and 3$g$. (Super-)critical pressures which have a ratio of unity at higher values of $S$ not shown.}
  \label{effectgrav}
\end{figure}

Fig. \ref{effectgrav} shows the ratio between calculated CO$_2$ pressures at 1$g$ and 3$g$.  At low stellar insolation, hence low CO$_2$ pressures (see Fig. \ref{maxpco2}), increasing gravity leads to cooler surface temperatures, and consequently lower CO$_2$ partial pressures (i.e., a ratio higher than 1 for all stars except the F star in Fig. \ref{effectgrav}). This indicates that the impact of the reduced GHE is dominating, in agreement with other studies of optically rather thin planetary atmospheres (e.g., \citealp{rauer2011}). In contrast, at higher stellar insolation (and correspondingly higher CO$_{2}$ pressures),  increasing gravity leads to warmer surface temperatures, hence higher CO$_2$ partial pressures  (i.e., a ratio lower than 1 in Fig. \ref{effectgrav}), implying that the decrease of the GHE is compensated by the decrease in planetary albedo. The influence of the stellar type is clearly seen in Fig. \ref{effectgrav}. For the M-star planet, with very little radiation in the Rayleigh regime (see Fig. \ref{energybalance}), the effect of increasing gravity is much higher than for the F-star planet, for which Rayleigh scattering is very important.


\label{discuss}	

\subsection{Implications for habitability}

\label{hz}

As can be inferred from Fig. \ref{maxpco2}, our calculations imply that relatively massive CO$_2$ atmospheres of the order of several bars are possible for almost all scenarios, even for planets orbiting far from their central star (stellar insolation $S$$\gtrsim$0.25).

At the triple point temperature of water, i.e. 273\,K, which permits its liquid phase, the CO$_2$ vapor pressure is about 34\,bar (see Fig. \ref{co2phase}). Hence, Fig. \ref{maxpco2} implies that liquid surface water can be achieved for stellar insolation $S_{\rm{34bar}}$ as low as $S_{\rm{34bar}}$=0.25 for the M-type star and $S_{\rm{34bar}}$=0.4 for the F-type star, providing a sufficiently large source of CO$_{2}$ is available for outgassing on the planet. This is, however, not the outer edge of the HZ, since surface temperature is not necessarily a monotonic function of CO$_{2}$ partial pressure (known as the maximum greenhouse effect, e.g., \citealp{kasting1993}). The CO$_{2}$ pressures corresponding to the maximum surface temperatures are therefore expected to be somewhat lower than the maximum CO$_{2}$ pressures in Fig. \ref{maxpco2}. Hence, the outer edge of the HZ is most likely located at lower stellar insolation (i.e., farther away from the star), than $S_{\rm{34bar}}$.

\subsection{N$_2$ partial pressure}

\begin{figure}[h]
  \resizebox{\hsize}{!}{ \includegraphics*{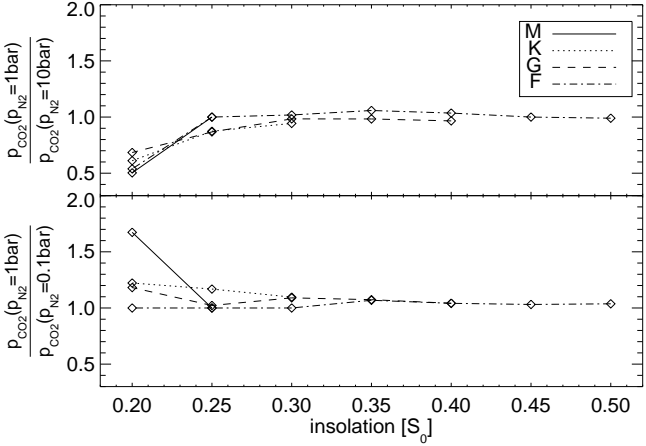}}\\
  \caption{Effect of varying N$_{2}$ partial pressure $p_{N_2}$ on the calculated maximum CO$_2$ pressures. (Super-)critical pressures which have a ratio of unity at higher values of $S$ not shown.}
  \label{effectn2}
\end{figure}

 The results of the N$_{2}$ study  (Sect. \ref{para} and Table \ref{planpar}) are shown in Fig. \ref{effectn2}.  As expected, for the high CO$_{2}$ partial pressures found for higher stellar insolation, the effect of varying N$_{2}$ is negligible, given that CO$_{2}$ is a much more efficient Rayleigh scatterer than N$_{2}$. However, for lower stellar insolation, and consequently lower CO$_{2}$ partial pressures, the effect of N$_{2}$ becomes discernible.

At these lower stellar insolation, the warming effect of adding N$_2$ to the atmosphere is clearly dominating, since the calculated maximum CO$_{2}$ pressures increase with increasing N$_{2}$ partial pressure. The effect is rather pronounced (almost a factor of 4 when increasing $p_{N_{2}}$ from 0.1 to 10\,bar) for the M star since Rayleigh scattering does not contribute greatly to the overall energy budget for these cases (most of the stellar radiation is emitted at wavelengths where Rayleigh scattering is negligible, see Fig. \ref{energybalance}). For the F-star simulations, maximum CO$_{2}$ pressures increase only by about 30\%, i.e. warming and cooling effects approximately cancel out.

\subsection{Surface albedo}

\begin{figure}[h]
    \resizebox{\hsize}{!}{\includegraphics*{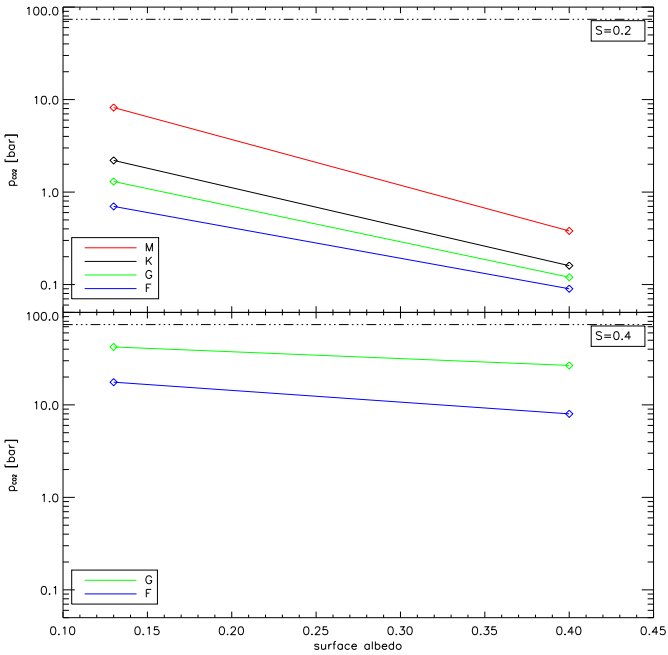}}\\
  \caption{Effect of varying surface albedo on the calculated maximum CO$_2$ pressures. (Super-)critical pressures at $S$=0.4 for the M and K star not shown (indicated by horizontal line at $p_{CO2}$=73.8\,bar).}
  \label{effectalb2}
\end{figure}

The results of the surface albedo study (Sect. \ref{para} and Table \ref{planpar}) are presented in Fig. \ref{effectalb2} which shows the decrease in calculated maximum CO$_2$ pressure when increasing the surface albedo.  At $S$=0.2, the decrease of CO$_2$ pressure is rather large, reaching about a factor of 20 for the M-type star. For a planet orbiting around an F-star, calculations imply maximum CO$_{2}$ pressures of the order of 0.1\,bar, so rather a teneous atmosphere. At $S$=0.4, the effect of increasing surface albedo is smaller than at $S$=0.2, but still reaches about a factor of 2-3 for the F-type star.

Fig. \ref{effectalb2} shows that, at $S$=0.2, the M-star planet is much more sensitive to a change in surface albedo (a reduction of a factor of about 20 in CO$_{2}$ pressure) than the F-star planet (a factor of 8), as seen by the steeper slope of the M-star line. The sensitivity is generally increasing for increasing stellar effective temperature (type from M to F). This is due to the larger amount of stellar energy emitted in the window regime (see Sect. \ref{stellar_type} and Fig. \ref{energybalance}).  Hence, the response to an increase in surface albedo, which affects principally the window, is more pronounced for the M-star planet and for lower stellar insolation (and correspondingly lower CO$_{2}$ partial pressures). For example, at $S$=0.4, the reduction for the F-star planet is decreased to about a factor of 2.

In order to investigate the combined effect of, e.g., an increase in N$_2$ partial pressure and an increase in surface albedo, we performed some additional test runs with both parameters changed. For the M star case, for example, the effect of N$_2$ was nearly unaltered even at high surface albedo. At $S$=0.2, an increase in surface albedo reduced the maximum CO$_2$ pressure from 8.2 to roughly 0.4\,bar (see Fig. \ref{effectalb2}) whereas an increase of N$_2$ partial pressure increased the maximum CO$_2$ from 8.2 to 16.3\,bar (see Fig. \ref{effectn2}). At high surface albedo and high N$_2$ pressure, the maximum CO$_2$ pressure obtained was 14.3\,bar, i.e. nearly as high as for the simulations at low surface albedo.

\section{Discussion}

\label{discuss}

\subsection{H$_{2}$O-CO$_{2}$ oceans}

As has been shown above (Fig. \ref{maxpco2}), for planets orbiting within the HZ of K-G-F stars there is a region of liquid surface CO$_{2}$ combined with surface temperatures above 273\,K, i.e. liquid surface H$_{2}$O. This means that it is possible to form H$_{2}$O-CO$_{2}$ oceans. Then, the question of planetary habitability would depend strongly on the pH of the liquid, even though extremophiles on Earth could support quite low pH values (e.g., \citealp{rothschild2001}). A detailed investigation of this interesting issue is however beyond the scope of this work.

\subsection{Implications of model assumptions }

The 1D atmospheric model used in this work is based on relatively few, simple assumptions. Most of these assumptions are physically justified, i.e. the assumption of adiabatic temperature gradients in the troposphere or radiative transfer as the main energy transport mechanism in the upper atmosphere. However, some of them (presence of clouds, greenhouse gases, water profile, etc.) are model-specific, hence need to be discussed further with respect to their possible influence on the results presented above.

The model is a cloud-free code, hence the potential impact of CO$_2$ clouds on the climate is neglected. It was shown by several authors that this potential impact could be quite large (e.g., \citealp{forget1997}, \citealp{mischna2000}, \citealp{cola2003},  \citealp{wordsworth2010}, \citealp{wordsworth2011}). However, this effect depends sensitively on cloud opacity, cloud coverage and cloud altitude. In addition, the effect of clouds is also probably very dependent on stellar type (see, e.g., \citealp{kitzmann2010} investigating the effect of stellar type for H$_{2}$O clouds). Investigating this is therefore a subject of further studies.

Furthermore, the model atmospheres considered in this work contained only the greenhouse gases CO$_2$ and water. This choice may be restrictive when applied to our own solar system, since other species, such as O$_{3}$, SO$_2$, CH$_4$, and N$_2$O, have been considered in models of the early Earth or early Mars climate (e.g., \citealp{yung1997}, \citealp{buick2007}, \citealp{haqq2009}). But given that the concentration of these gases depend on very specific planetary scenarios (e.g., outgassing history, biospheric evolution, etc.), assuming them in the context of exoplanets (without any geological or other constraints) is rather arbitrary. However, the impact on stratospheric temperatures through the absorption of UV (e.g., O$_{3}$ and SO$_2$) or near-IR (e.g., CH$_{4}$) stellar radiation is potentially important.

Radiative transfer in dense, CO$_{2}$-dominated atmospheres presents many challenges (e.g., collision-induced absorption, sub-Lorentzian behavior of line wings, etc.). The parametrization of the collision-induced absorption (CIA) used in this study is taken from \citet{kasting1984}. A recent study \citep{wordsworth2010cont} presented a revised parametrization, showing that the calculation presented by  \citet{kasting1984} most likely over-estimates the opacity. In order to estimate the impact of the CIA  uncertainties on our results, we performed a sensitivity study with a reduced (by roughly a factor of 2) CIA. The conclusions however did not change qualitatively. At $S$=0.2, calculated CO$_2$ maximum pressures around K-, G- and F-stars decreased by less than 50\%, for the M-star the maximum CO$_2$ pressure decreased from 8.2 to 3.0\,bar. At $S$=0.35, results changed less than 20\% except for the K-star planet, where a maximum CO$_2$ pressure of 47.3\,bar was calculated, instead of 73.8\,bar (i.e., the critical pressure of CO$_2$, see Fig. \ref{maxpco2}). Therefore, our calculations (using \citealp{kasting1984}) are likely to be overestimates of the maximum CO$_2$ partial pressures. 

The model uses a super-saturation of $\mathcal{S}_s$=1.34 to determine the CO$_{2}$ convective regime (see eq. \ref{co2glan}). The choice of $\mathcal{S}_s$ has been shown to be very important for early Mars climate simulations, e.g. \citet{poll1987} (using $\mathcal{S}_s$=$\infty$) find significantly higher surface temperatures ($>$30\,K) than \citet{kasting1991} (using $\mathcal{S}_s$=1).  The assumed $\mathcal{S}_s$=1.34 is based on \citet{glandorf2002}, a value observed for specific conditions (e.g., dust loading available for nucleation) which could be different on exoplanets (as low as  $\mathcal{S}_s$=1, but also possibly significantly higher). In this sense, the calculated maximum CO$_{2}$ pressures are not necessarily upper limits. To further investigate this, we performed some sensitivity simulations with  $\mathcal{S}_s$=1. As expected, calculated maximum CO$_2$ pressures were lower, of the same order of magnitude as for the CIA study mentioned above. However, the main conclusions obtained in this work (i.e., the existence of maximum CO$_2$ pressures far below the critical pressure) were not affected.

The relative humidity profile used in this work \citep{manabewetherald1967} has been derived from observations of modern Earth. It has been used in many 1D simulations of terrestrial exoplanets, both Earth-like (e.g., \citealp{Seg2003}, \citealp{Grenf2007asbio}) and not (e.g., \citealp{vparis2010gliese}, \citealp{wordsworth2010}).  Since the humidity profile is anything but trivial to model in 1D simulations, some authors chose to fix relative humidity at an isoprofile (e.g., \citealp{kasting1991}). However, given the large amounts of CO$_{2}$ in the model atmospheres (73.8\,bar at 303\,K), the impact of water (42\,mbar at 303\,K) on atmospheric structure (via near-IR absorption) and surface conditions (via the GHE) is somewhat negligible. Therefore, the choice of the relative humidity profile is probably not important.

\subsection{Synchronous rotation}

For planets orbiting very close to their star, tidal locking of the planetary rotation with the orbital period is very likely. The time scale $t_{\rm{lock}}$ of tidal locking is very sensitive to orbital distance ($t_{\rm{lock}}\sim a^{6}$, $a$ orbital distance, see e.g. \citealp{griess2005}). Hence, tidal locking is mainly an issue for the habitability of planets orbiting around M stars due to the closeness of the HZ to the star. It has been argued that for planets with a perpetual nightside, the atmosphere could collapse since the nightside forms a cold trap for the volatiles, which, in the context of this work, could present an alternative way of obtaining maximum CO$_{2}$ pressures. However, as has been shown by numerous modeling studies (e.g., \citealp{joshi1997}, \citealp{joshi2003}, \citealp{wordsworth2011}, \citealp{kite2011}), moderately dense atmospheres containing hundreds of millibars or more of CO$_{2}$ are sufficient to avoid atmospheric collapse by means of atmospheric circulation. Hence, the M-star simulations presented in this work are not thought to be subject to atmospheric collapse induced by synchronous rotation.

\section{Conclusions}

\label{summary}

We have presented a detailed parameter study to constrain the maximum CO$_2$ partial pressure possible for terrestrial exoplanets, using a 1D cloud-free atmospheric model. Parameters investigated included the central star type, the orbital distance and the planetary gravity. Furthermore, we investigated the influence of N$_{2}$ partial pressure and the surface albedo on the maximum CO$_2$ partial pressure.  

Results imply that super-critical atmospheres (i.e., $p_{\rm{CO}_{2}}$$\geq$$p_{\rm{crit}}=$73.8\,bar) are possible for planets around M stars for stellar insolation corresponding to $S_{\rm{crit}}$=0.25 or higher. For increasingly bluer stars (i.e., higher effective temperatures), this super-critical stellar insolation increases (e.g.,  $S_{\rm{crit}}$$>$0.5 for an F-type star). For lower stellar insolation, the calculations presented here imply that there is indeed a maximum CO$_2$ partial pressure, even if the planets are orbiting well within the habitable zone. Nevertheless, massive CO$_{2}$ atmospheres of the order of bars are still possible for most scenarios. For planets orbiting very far from an F-type central star (e.g., $S$=0.2 in this work), CO$_2$ partial pressures could be constrained to be less than 1\,bar.

The effect of planetary gravity is twofold. For low stellar insolation and corresponding cold surface temperatures, increasing planetary gravity leads to a decrease of maximum CO$_2$ partial pressure due to less atmospheric greenhouse effect. At higher stellar insolation, an increase of planetary gravity increases the calculated maximum CO$_2$ partial pressure because of less Rayleigh scattering. For F-star planets, the effect is up to a factor of 2, depending on stellar insolation, whereas for an M-star planet, a factor of about 4 has been calculated.

Increasing the N$_{2}$ partial pressure leads to warmer surface temperatures for all cases, hence corresponding maximum CO$_{2}$ pressures are higher. The effect reaches up to a factor of 3 for planets around an M star, upon increasing the N$_{2}$ partial pressure from 0.1 to 10\,bar. The surface albedo has an important effect on the values of the maximum CO$_2$ partial pressure. A higher surface albedo leads to cooler surface temperatures, hence less CO$_{2}$ in the atmosphere. Decreases of about a factor of 20 have been shown when increasing the surface albedo from 0.13 to 0.4.

The presence of CO$_2$ and H$_2$O clouds could alter these results because of their potentially large impact on the planetary energy balance. However, our (clear-sky) results show in a robust way that the composition and the evolution of planetary atmospheres strongly depend on orbital and planetary parameters. Although a consistent model for the determination of CO$_{2}$ partial pressures must take processes such as sequestration or outgassing into account, our results show that there is a fundamental thermodynamic limit to the amount of CO$_{2}$ in terrestrial atmospheres, independent of the planetary reservoir. Hence, a more detailed coupling between interior, surface and atmosphere models should be used to accurately predict atmospheric composition of terrestrial planets.

\begin{acknowledgements}

P.v.P., P.H. and F. Selsis acknowledge support from the European Research
Council (Starting Grant 209622: E$_3$ARTHs). This research has been partly supported by the Helmholtz Association through the research alliance "Planetary Evolution and Life". We thank M. Godolt and D. Kitzmann for valuable discussions and comments regarding this manuscript.

\end{acknowledgements}

\bibliographystyle{aa}
\bibliography{literatur_co2_max}

\end{document}